\documentclass[aps,prb,onecolumn,amsmath,amssymb,superscriptaddress,floatfix,12pt]{revtex4}

\usepackage{multirow}
\usepackage{xcolor}
\usepackage{bm}
\usepackage{amsfonts,amssymb,amsmath}
\usepackage{graphicx,dcolumn,bm,color,braket,slashed}
\usepackage{times} 
\usepackage{comment}
\usepackage{array}
\usepackage{textcomp}
\usepackage{siunitx}
\usepackage{hyperref}

\begin{document}
\title{Engineering two-dimensional nodal semimetals in functionalized biphenylene by fluorine adatoms}

\author{Seongjun Mo}
\altaffiliation{These authors contributed equally to this work}
\affiliation{Department of Physics, Konkuk University, Seoul 05029, Korea}

\author{Jaeuk Seo}
\altaffiliation{These authors contributed equally to this work}
\affiliation{Department of Physics, Ajou University, Suwon 16499, Korea}
\affiliation{Department of Physics, Korea Advanced Institute of Science and Technology, Daejeon 34141, Korea}

\author{Seok-Kyun Son }
\affiliation{Department of Physics, Kyung Hee University, Seoul 02447, Republic of Korea }
\affiliation{Department of Information Display, Kyung Hee University, Seoul 02447, Republic of Korea }

\author{Sejoong Kim}
\email{sejoong@alum.mit.edu}
\affiliation{University of Science and Technology (UST), Gajeong-ro 217, Daejeon 34113, Korea}
\affiliation{Korea Institute for Advanced Study, Hoegiro 85, Seoul 02455, Korea}

\author{Jun-Won Rhim}
\email{jwrhim@ajou.ac.kr}
\affiliation{Department of Physics, Ajou University, Suwon 16499, Korea}
\affiliation{Research Center for Novel Epitaxial Quantum Architectures, Department of Physics, Seoul National University, Seoul, 08826, Korea}

\author{Hoonkyung Lee}
\email{hkiee3@konkuk.ac.kr}
\affiliation{Department of Physics, Konkuk University, Seoul 05029, Korea}
\affiliation{Research Center for Novel Epitaxial Quantum Architectures, Department of Physics, Seoul National University, Seoul, 08826, Korea}

\begin{abstract}
We propose a new band engineering scheme on the biphenylene network, a newly synthesized carbon allotrope.
First, we investigate the mechanism for the appearance of type II Dirac fermion in a pristine biphenylene network.
We show that the essential ingredients are mirror symmetries and the stabilization of the compact localized eigenstates via destructive interference.
While the former is used for the band-crossing point along high symmetry lines, the latter makes the obtained Dirac dispersion highly inclined.
Then, we demonstrate that many other different kinds of Dirac fermions, such as type-I Dirac, gapped type-II Dirac, and nodal line semimetals, can be developed by fluorinating the biphenylene network periodically in various ways.
In this program, the key role of the fluorine atoms is manipulating the condition of the destructive interference and mirror symmetries.
\end{abstract}

\maketitle

\section{Introduction}
Triggered by the discovery of carbon nanotubes~\cite{Iijima1991, DRESSELHAUS1995883}, studies on low dimensional carbon allotropes, such as graphene, have significantly proliferated because one can have a variety of intriguing electronic structures depending on the arrangement of carbon atoms. 
%
For example, one can have massless Dirac fermions in graphene~\cite{novoselov2004electric,zhang2005experimental}, whereas infinitely heavy particles can also appear in another honeycomb network of carbon atoms called the cyclic-graphdiyne, which hosts a singular flat band~\cite{You2019,rhim2020quantum}.
In the case of graphene, relying on its high structural stability, one can further engineer the band structure to obtain flat bands by tailoring it into ribbon geometries~\cite{PhysRevLett.97.216803} or twisting two stacked graphene sheets~\cite{Cao2018, PhysRevLett.122.106405}.
These peculiar electronic structures have received great attention because they are relevant to the possible many-body phases such as ferromagnetism~\cite{mielke1993ferromagnetism,tasaki1998nagaoka} and superconductivity~\cite{volovik1994fermi,balents2020superconductivity,peri2021fragile,volovik2018graphite}.

Recently, another type of two-dimensional carbon allotrope, called the biphenylene network (BPN), was synthesized~\cite{doi:10.1126/science.abg4509.Fan} and has attracted significant interests in various properties of BPN layers including electronic, optical, mechanical, thermal, magnetic and chemical properties~\cite{doi:10.1021/acs.nanolett.2c00528.Son, PhysRevB.104.235422, PhysRevB.105.035408, Luo2021, https://doi.org/10.1002/jcc.26854, LIU2022153993, CHOWDHURY2022110909, D1CP04481H, D2RA03673H, 10.1063/5.0102085, HAMEDMASHHADZADEH2022111761, MORTAZAVI2022100347, D1NR07959J, REN2023112119, Bafekry_2022, 10.1063/5.0088033, Zhang_2022, doi:10.1021/jacs.2c02178, Ren_2022, Ge_2022, D2CP04381E, VEERAVENKATA2021893, D1TC04154A, https://doi.org/10.1002/aenm.202200657, LI2022349, XIE2023112041, 10.1063/5.0140014, doi:10.1021/acs.jpclett.1c03851, https://doi.org/10.1002/est2.377, Asadi2022, SU2022108897, Al-Jayyousi2022, D2CP00798C, D2CP04752G, D2NH00528J, PhysRevB.107.045422, Yang_2023}.
Moreover, from the first principles analysis, BPN turned out to exhibit another intriguing band dispersion called type-II Dirac fermion consisting of two heavily inclined cones, so that we have open Fermi surfaces instead of Fermi points or circles~\cite{doi:10.1021/acs.nanolett.2c00528.Son, PhysRevB.104.235422}.
In type-II Dirac semimetals~\cite{RevModPhys.90.015001.2018}, one can have electron- and hole-type carries simultaneously in contrast to the type-I case such as graphene.
This intricate band shape may lead to a variety of unusual electronic phenomena such as anisotropic transport and magnetoresistance behavior~\cite{wang2016gate,chen2016extremely,kumar2017extremely,lai2018broadband} and undamped gapless plasmon modes~\cite{sadhukhan2020novel}.
Therefore, engineering their band shapes is important to tune the electronic properties for applications.

\begin{figure}[t]
\centering
\includegraphics[width=0.65\textwidth ]{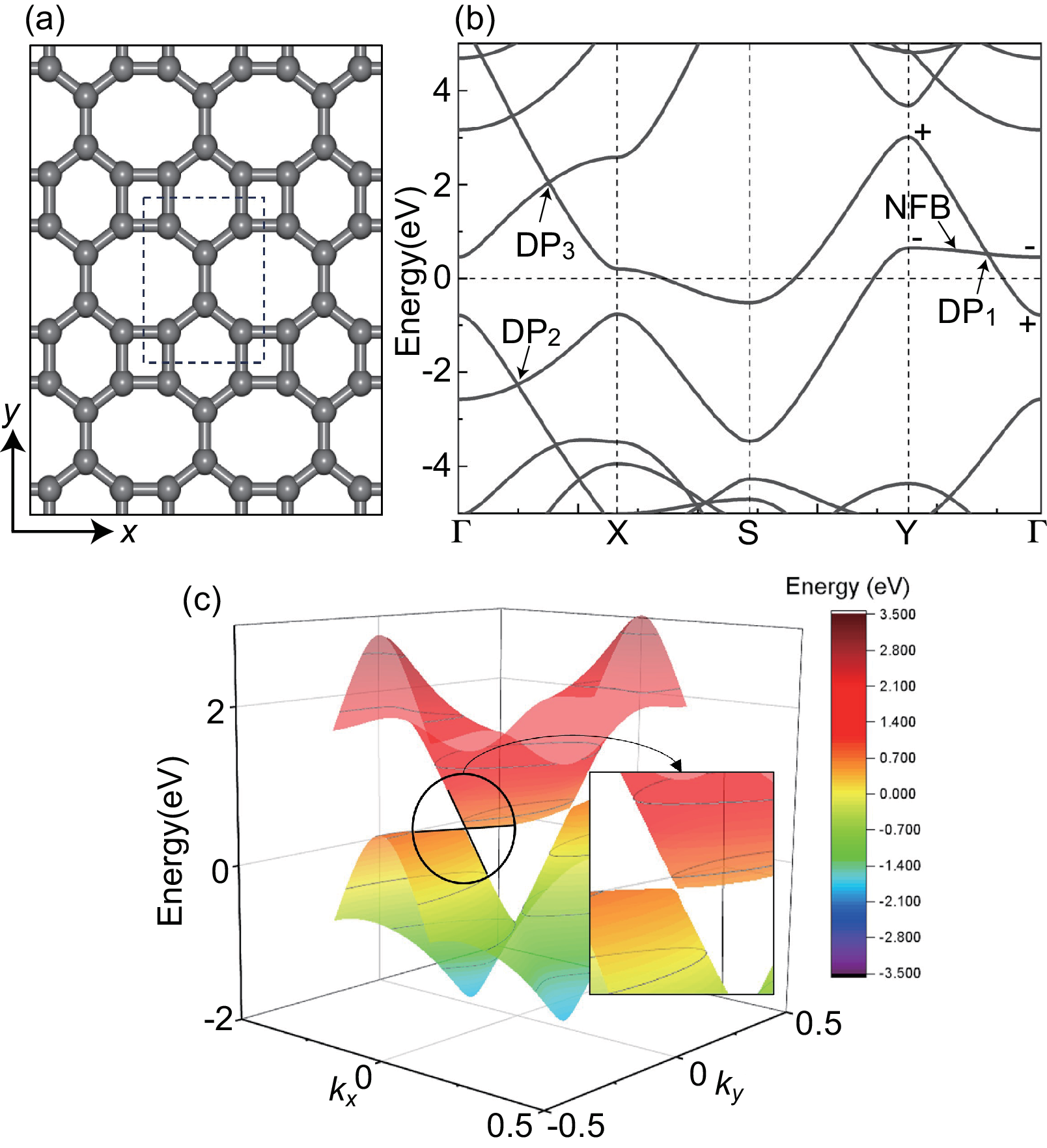}
\caption{(a) The lattice structure, (b) band structure, and (c) 3D energy structure near the $\Gamma$Y line of pristine BPN. The unit cell is represented by a dashed box in (a). Several Dirac points are denoted by DP$_1$, DP$_2$, and DP$_3$, in (b). NFB indicates the nearly flat band.
}\label{fig:pristine_dft}
\end{figure}

%
In this paper, we first analyze the origin of the type-II Dirac dispersion of the pristine BPN from the perspective of destructive interference and symmetry.
The essential feature of the type-II Dirac dispersion of the pristine BPN is that it is inclined so that a flat band with a zero Fermi velocity appears along a high symmetry line while the Dirac band-crossing is protected by mirror symmetry.
We show that the the pristine BPN hosts a proper destructive interference stabilizing a stripe-type compact localized state, which signals the existence of a flat band, along the mirror-symmetric axis in momentum space.
This implies that one can engineer the electronic structures of the BPN by controlling the condition for destructive interference and symmetries.
%
We demonstrate that this can be done successfully by absorbing fluorine atoms at various positions.
We show that we can have diverse Dirac dispersions of different kinds, such as type-I, massive, and nodal line Dirac fermions, by fluorinating BPN periodically.

\section{Results and Discussion}
\subsection{First-principles calculations}
To investigate the geometrical and electronic properties of pristine single-layer BPN, we perform geometry optimization of BPN using first-principles calculations based on density functional theory (DFT). 
The optimized geometry and energy bands of BPN are shown in Fig.~\ref{fig:pristine_dft}(a) and (b), respectively. 
As in the previous study~\cite{doi:10.1021/acs.nanolett.2c00528.Son, PhysRevB.104.235422}, we observe a couple of type-II Dirac points whose Dirac nodes lie on $\Gamma$Y near the Fermi energy $E_{F}$ as denoted by DP$_1$ in Fig.~\ref{fig:pristine_dft}(b). 
As highlighted in Fig.~\ref{fig:pristine_dft}(c), there are two tilted Dirac dispersions around the $\Gamma$ point. corresponding to DP$_1$ in Fig.~\ref{fig:pristine_dft}(b). 
While the Dirac point is developed by the crossing of two bands along $\Gamma$Y, it is essential that one of them is almost flat for realizing a type-II Dirac fermion.
Therefore, it is crucial to understand the origin of the flat dispersion along $\Gamma$Y to reveal the mechanism for the appearance of the type-II Dirac semimetal, and it will be discussed in the following subsection.

\begin{figure}[t]
\centering
\includegraphics[width=1\textwidth ]{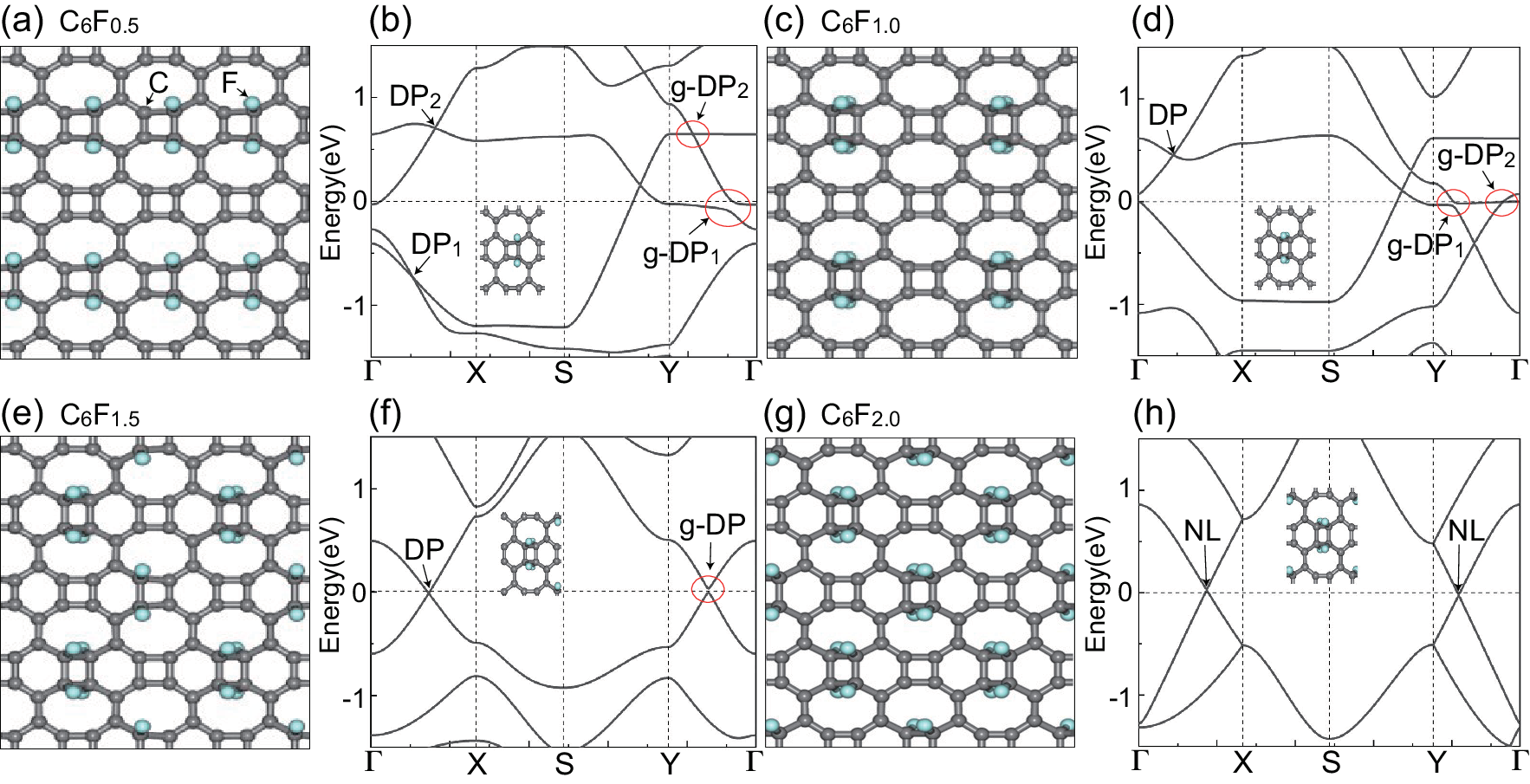}
\caption{Optimized geometries of fluorinated BPN monolayers with different concentrations of fluorine and their electronic structures: (a,b) C$_6$F$_{0.5}$, (c,d) C$_6$F$_{1.0}$, (e,f) C$_6$F$_{1.5}$, and (g,h) C$_6$F$_{2.0}$. Gray (Cyon) spheres represent carbon (fluorine) atoms. Dirac nodes, gapped-Dirac points, and nodal lines are denoted by DP, g-DP, and NL, respectively. 
}\label{fig:fbpn_lattice}
\end{figure}

We introduce fluorine atoms to a single-layer BPN to engineer its electronic structures.
Regarding a pristine BPN monolayer, binding energies of fluorine (F) atoms to carbon sites are computed as follows: 3.20 eV/F for square, 2.21 eV/F for hexagon, and 2.10 eV/F for octagon. 
It implies that F atoms energetically prefer to bind to a square unit rather than hexagon and octagon units. 
%
Varying fluorine concentrations $x$, we consider four fluorinated BPN layers C$_6$F$_{x}$ for $x=$ 0.5, 1.0, 1.5, and 2.0, 
whose atomic structures are displayed in Figs.~\ref{fig:fbpn_lattice}(a), (c), (e), and (g), respectively.
%
%
DFT calculations also reveal that F atoms energetically prefer to attach to square sites for higher F concentrations. 
Corresponding binding energies of F to square units are 3.45, 3.51, 3.48, and 3.50 eV/F for $x=$ 0.5, 1.0, 1.5, and 2.0, respectively. 
Note that due to $sp^3$ hybridization the bond between C and F is not perpendicular to single layer BPN.

Band structures of the fluorinated BPNs are plotted on the right side of atomic configurations in Fig.~\ref{fig:fbpn_lattice}
Intriguingly, diverse nodal states emerge depending on F concentrations. 
At fluorine concentrations of 0.5 and 1.0, gapped type-II Weyl nodes are found as a result of crossing between flat bands near $E_{F}$ along $\Gamma$Y and energy bands hosting saddle-point van Hove singularity (vHS) at $Y$. 
See red circles in Figs.~\ref{fig:fbpn_lattice}(b) and (d). 
These gap opening is attributed to the mirror symmetry breaking with respect to monolayer BPN plane due to fluorine attachment.
On the other hand, we found gapless and gapped type-I Dirac points along $x$- and $y$-axis in C$_6$F$_{1.5}$ as shown in Fig.~\ref{fig:fbpn_lattice}(f) while nodal rings appear in C$_6$F$_{2}$ as noted in Fig.~\ref{fig:fbpn_lattice}(h).
For clarity, we provide 3D band structures of these two compounds in Supplementary Fig. 3.
%
An effective Hamiltonian is formulated using a tight-binding model in the next subsection to provide a more accurate interpretation.

To understand how the variety of relativistic dispersions can be stabilized in pristine BPN and fluorinated BPNs, we conduct an analysis of their wave function symmetries.
Here, we first focus on protection or gap-opening mechanisms of the nodal points or lines, while the origin of the inclination in Dirac dispersions is addressed in the next subsection.
Specifically, we only consider the wave functions of the bands near the Fermi level. 
The symmetries of the wave functions are visualized by plotting the amplitudes of the Bloch wave function at $\Gamma$ point as illustrated in Supplementary Fig.~1.
For the pristine BPN, one can show that the band-crossing along $\Gamma$Y in Fig.~\ref{fig:pristine_dft}(b) is protected by the mirror symmetry with respect to $yz$ plane, which is denoted by $M_{x}$.
The nearly flat band (NFB) and the dispersive band, which constitute the type-II Dirac point, correspond to the mirror eigenvalue -1 and 1, respectively, as shown in Supplementary Fig.~1(a).
Furthermore, upon close inspection of the wave function depicted in Supplementary Figure 1(a), an intriguing resemblance to the anti-bond wave function of benzene becomes apparent. 
This finding may imply a potential connection between the electronic properties of the material under investigation and those of benzene, which is known for its aromatic properties and unique bonding characteristics.

On the other hand, the mirror symmetry $M_{x}$ is broken in C$_6$F$_{0.5}$ and C$_6$F$_{1}$, 
As a result, the wave functions are not mirror-symmetric with respect to $yz$ plane as shown in Supplementary Figure 1(b) and (c), and band crossings along $\Gamma$Y are all gapped out, realizing massive type-II Dirac fermions as shown in Figs.~\ref{fig:fbpn_lattice}(b) and (d).
Although several type-II Dirac nodes, denoted by g-DP$_2$ in Fig.~\ref{fig:fbpn_lattice}(b) and (d), looks gapless, they actually have tiny gaps.
Note that C$_6$F$_{0.5}$ and C$_6$F$_{1}$ respect the mirror symmetry $M_{y}$ with respect to $zx$ plane, so their wave functions are either mirror symmetric or mirror anti-symmetric as illustrated in as illustrated in Supplementary Figure 1(b) and (c).
Thus band-crossings indicated by DP$_1$ and DP$_2$ in Fig.~\ref{fig:fbpn_lattice}(b) and DP in Fig.~\ref{fig:fbpn_lattice}(d), are protected along $\Gamma$X.
As like the two previous compounds, C$_6$F$_{1.5}$ satisfies only the mirror symmetry $M_{y}$, so it has gapless and gapped-out Dirac nodes on $\Gamma$X and $\Gamma$Y, respectively. See Fig.~\ref{fig:fbpn_lattice}(f).
In contrast to C$_6$F$_{0.5}$ and C$_6$F$_{1}$, C$_6$F$_{1.5}$ does not stabilize flat bands, resulting exclusively in type-I Dirac fermions.
Finally, in C$_6$F$_{2}$, both mirror symmetries $M_{x}$ and $M_{y}$ are respected.
The system, therefore, becomes gapless along both $\Gamma$X and $\Gamma$Y as illustrated in Fig.~\ref{fig:fbpn_lattice}(h).
In fact, these two nodes are part of the nodal line.

\subsection{Tight-binding analysis}
In this subsection, we discuss another crucial condition for the appearance of type-II Dirac fermions in pristine BPN and aforementioned fluorinated BPNs by analyzing how flat bands are stabilized along certain symmetry axes.
%
%
To this end, we apply a tight-binding method, which is advantageous for the flat band analysis.
The essential mechanism for the development of the dispersionless band is the existence of a compact localized state(CLS), which is a localized eigenstate having nonzero amplitudes only inside a finite region~\cite{bergman2008band,leykam2018artificial,rhim2019classification,ma2020direct}.
Although electrons can move on the lattice via hopping processes, such an extremely localized mode can exist due to the destructive interference hosted by the special lattice structures.
The CLS is considered a characteristic eigenstate of a flat band because it is guaranteed to exist when it is completely flat~\cite{rhim2019classification}.
While one can have $N$ (the number of unit cells) different CLSs centered at different positions, they are not independent of each other if the Bloch eigenstate corresponding to the flat band possesses a discontinuity in momentum space.
Such a flat band is called a singular flat band, and its geometric and topological aspects have been studied extensively~\cite{rhim2019classification,ma2020direct,rhim2020quantum,rhim2021singular,hwang2021geometric}.
%
%

\begin{figure}[t]
\centering
\includegraphics[width=0.8\textwidth ]{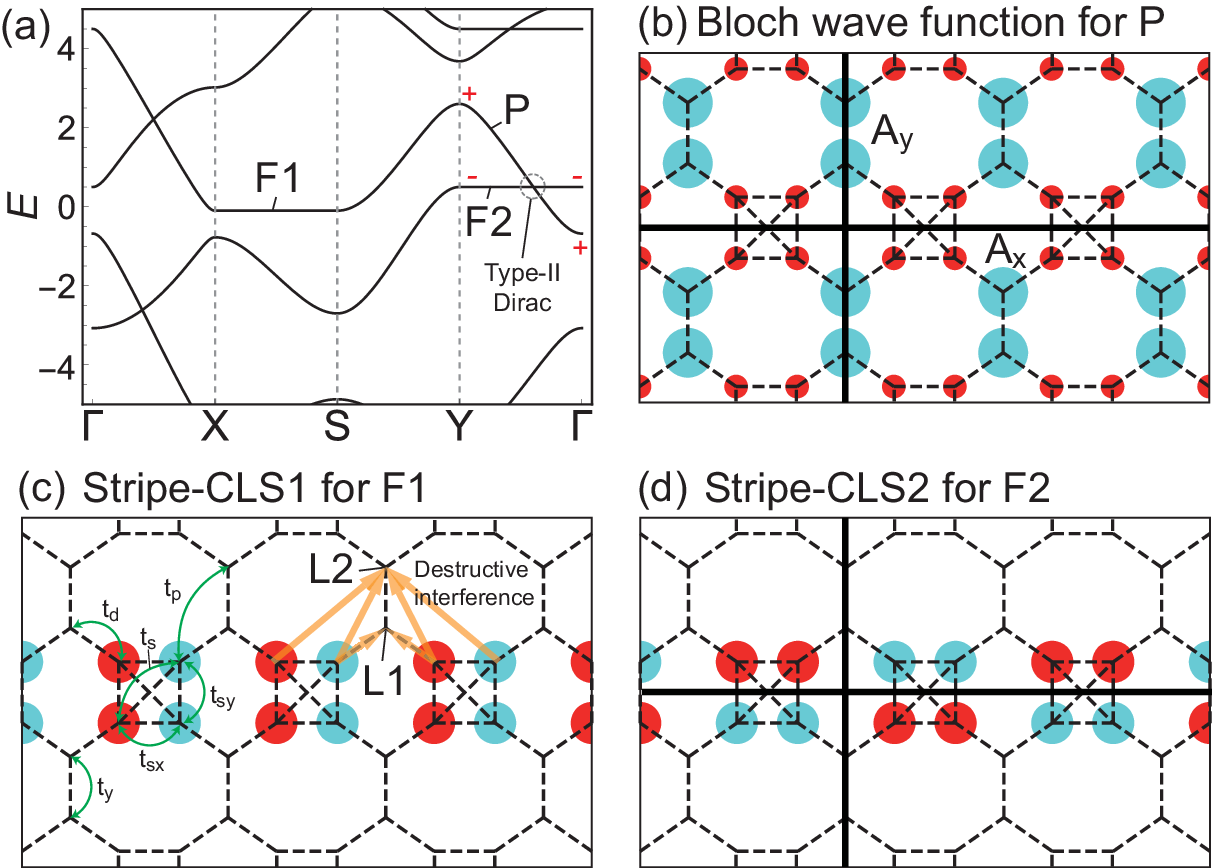}
\caption{(a) The tight-binding band structures of the pristine BPN with band parameters $\{ t_{sx},t_{sy},t_{y},t_d,t_s,t_p,E_0 \} = \{ -3,-2.7,-2.7,-2.8,-0.7,0,-0.5\}$. We indicate the 1D flat bands along XS and Y$\Gamma$ by F1 and F2, respectively. The parabolic band crossing with the flat band F2 is denoted by P. (b) The Bloch wave function corresponding to the parabolic band P at $\Gamma$. Colored circles represent the amplitudes of the wave function. Red (cyan) color implies that the sign of the amplitude is 1 (-1), while the radius of the circle is proportional to the magnitude of the amplitude. Thick vertical and horizontal lines represent the mirror symmetry planes. In (c) and (d), we illustrate the stripe-CLS1 and stripe-CLS2 corresponding to the flat bands F1 and F2, respectively.  In (c), the hopping processes are summarized. The CLS can be an eigenmode even if the longer-ranged hopping processes represented by the yellow arrow are further included due to the destructive interference at sites L1 and L2.
}\label{fig:pristine_tb}
\end{figure}

First, regarding that energy bands of our interest mostly originate from $p_{z}$ orbitals of carbon atoms, we construct an effective tight-binding model consisting of the six $p_{z}$ orbitals for the pristine BPN to understand the origin of its flat bands along $\Gamma$Y and XS, as shown in Fig.~\ref{fig:pristine_tb}(a).
Along these lines, the Hamiltonian can be regarded effectively as a one-dimensional system hosting flat bands.
The effective 1D Hamiltonian along $\Gamma$Y and XS are obtained by the inverse Fourier transform of the Bloch Hamiltonian with $k_x=0$ and $k_x=\pi$, respectively.
Therefore, we seek a CLS compactly localized along $y$-axis in order to understand the origin of flat bands.
%
%
Six tight-binding parameters are denoted by $t_{sx}$, $t_{sy}$, $t_y$, $t_d$, $t_s$, and $t_p$ as represented in Fig.~\ref{fig:pristine_tb}(c).
Main features of the DFT calculations can be captured by the tight-binding parameters $\{ t_{sx},t_{sy},t_{y},t_d,t_s,t_p,E_0 \} = \{ -3,-2.7,-2.7,-2.8,-0.7,-0.3,-0.5\}$, where $E_0$ is the overall energy shift.
Two perfectly flat bands around the Fermi level along $\Gamma$Y and XS are denoted by $F1$ and $F2$, respectively.
Their energies are given by $E_{F1} = -t_p - t_s + t_{sx} - t_{sy} + E_0$ and $E_{F2} = -t_p - t_s - t_{sx} + t_{sy} + E_0$.
The flatness of these flat bands is robust against the variation of those tight-binding parameters.
%
%
The CLSs corresponding to the flat bands F1 and F2 are plotted in Fig.~\ref{fig:pristine_tb}(c) and (d).
As the effective Hamiltonians are translationally invariant along $y$-axis, the CLSs are compactly localized along the same direction.
On the other hand, they are extended along $x$-direction modulating with the fixed momenta $k_x=0$ and $k_x=\pi$.
These CLSs are denoted as the stripe-CLSs.
One can notice that the lattice structure provides a destructive interference at the sites linking two neighboring square plaquettes of carbon atoms, denoted by $L_1$, as explained in Fig.~\ref{fig:pristine_tb}(c).
While the flat band $F2$ has a Dirac band-crossing with a dispersive band, this is protected by mirror symmetry $M_x$ because the CLS and the wave functions in the dispersive band have mirror eigenvalues $-1$ and $+1$, respectively, as shown in Fig.~\ref{fig:pristine_tb}(b) and (d).
This band-crossing between the flat and dispersive bands results in type-II Dirac fermion, as plotted in Fig.~\ref{fig:pristine_tb}(a).
Note that a long-range hopping $t_p$ does not break the flatness of the flat band because this hopping process also offers another destructive interference for the same CLS, denoted by $L_2$ in Fig.~\ref{fig:pristine_tb}(c).
Although the longer-ranged hybridizations would not provide such a destructive interference and, therefore, deform the flat band, the resultant bandwidth is tiny, as we observed in DFT calculations, because their hopping amplitudes should be much smaller than the nearest neighbor ones.
Even if the flat band is warped, the band-crossing is robust against the inclusion of the long-range processes because mirror symmetry is still respected.
The pristine BPN has another mirror symmetry $M_y$, and the type-I Dirac dispersions along $\Gamma$X are protected by it.

\begin{figure}[htb]
\centering
\includegraphics[width=0.8\textwidth ]{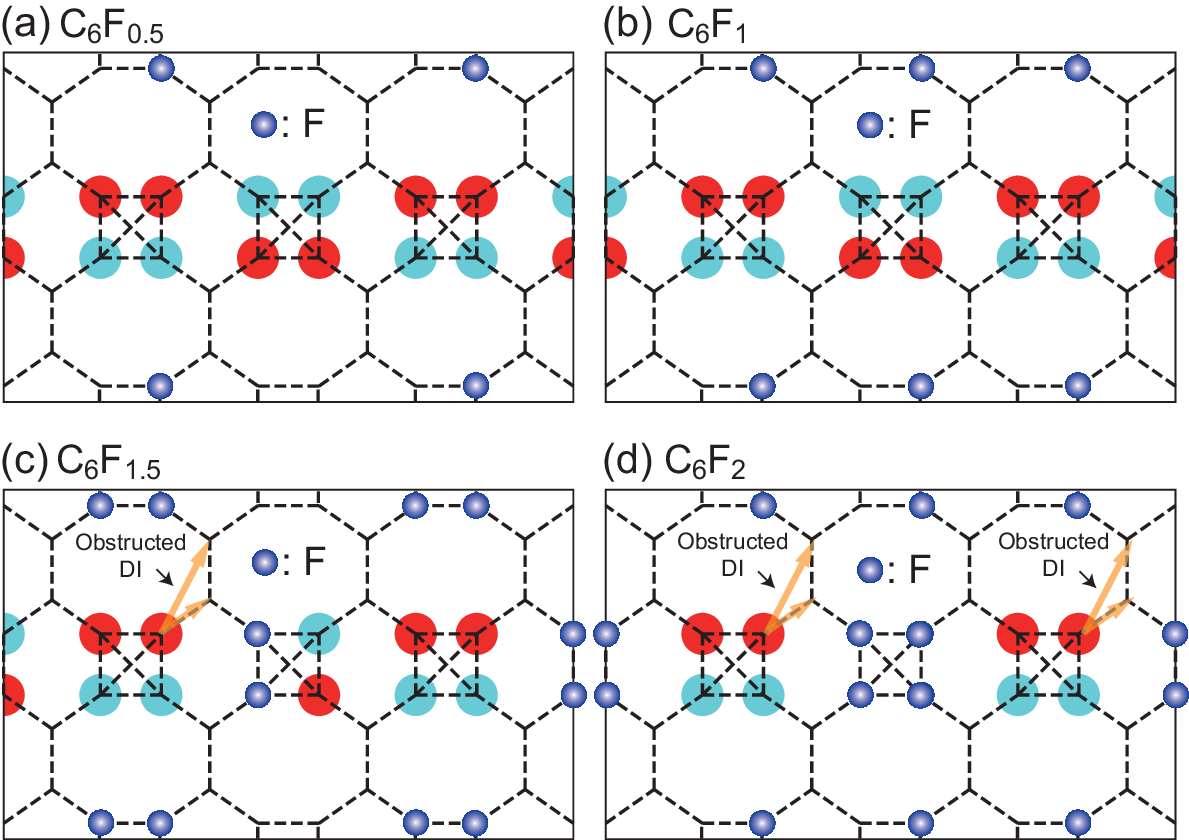}
\caption{In (a) and (b), the stripe-CLSs of C$_6$F$_{0.5}$ and C$_6$F$_{1}$ are plotted. They are occupying the chain of square plaquettes of carbon atoms in the middle. The attached fluorine atoms are represented by blue circles. The red and cyan circles mean plus and minus signs, respectively. In (c) and (d), we show that the stabilization of the stripe-CLSs fails due to the obstructed destructive interference(DI) by fluorine atoms in the middle chain of the square plaquettes.
}\label{fig:fluo_bp_tb}
\end{figure}

As a next step, we show that by fluorinating BPN, such as C$_6$F$_{0.5}$ and C$_6$F$_{1}$, we can have gapped type-II Weyl semimetals.
In the tight-binding analysis, the fluorination is assumed to be equivalent to making a vacancy at the corresponding carbon site. 
This assumption is reasonable since hybridization between $p_{z}$ orbital and fluorine atoms leads to bonding and anti-bonding states whose energies might be pushed away from the energy window of our interest. 
For C$_6$F$_{0.5}$ and C$_6$F$_{1}$, stripe-CLSs can still be stabilized along the chain of square plaquettes of carbon atoms, which are not fluorinated and extended along $x$-axis, as shown in Fig.~\ref{fig:fluo_bp_tb}(a) and (b).
As a result, the flat bands appear along $\Gamma$Y and XS, as shown by the nearly flat bands in DFT band structures in Fig.~\ref{fig:fbpn_lattice}(b) and (d).
However, due to the broken mirror symmetry by the fluorine atoms, any band-crossing along these high-symmetry lines does not have to be protected.
Indeed, all the band-crossings of type-II Dirac dispersions along $\Gamma$Y are gapped out, as discussed in the previous subsection.
In the case of C$_6$F$_{0.5}$, we note that the energy gap at the higher Dirac point (g-DP$_{2}$) is tiny but nonzero.
Namely, we obtained massive type-II Dirac fermions by fluorination, which breaks mirror symmetry while maintaining destructive interference.
Since the attached fluorine atoms do not break $M_y$, type-I Dirac dispersions along $\Gamma$X and SY are all immune from being gapped out.
If we attach more fluorine atoms, any stripe-CLSs cannot be an eigenmode because all the possible destructive interferences are obstructed by the fluorine atoms, as shown in Fig.~\ref{fig:fluo_bp_tb}(c) and (d).
Therefore, we cannot expect flat bands in C$_6$F$_{1.5}$ and C$_6$F$_{2}$, and no type-II Dirac semimetals are expected in these compounds.
Instead, type-I and nodal line semimetals appear, as shown in DFT calculations.
%
%
%
%
%
%

\section{Conclusion}
In this paper, we have shown that a variety of Dirac particles, such as massless or massive type-I and -II Dirac, and a nodal line can be obtained by the manipulated fluorination on the BPN.
In our band engineering scheme, it is crucial that one can eliminate destructive interferences and mirror symmetries by properly locating the fluorine atoms.
While we have focused on a specific material system, our work eventually proposes a novel band engineering method, where we control the slope of a part of Dirac dispersion by manipulating the condition for destructive interference via the molecular absorption technique.

Attaching atoms on two-dimensional lattices or surfaces on a microscopic level is experimentally feasible.
By using scanning tunneling microscopy(STM), hydrogen atoms or CO molecules can be absorbed and controlled on graphene or Cu(111) surface on the atomic scale~\cite{gonzalez2016atomic,slot2017experimental}, and even an automated manipulation of their position is possible~\cite{moller2017automated}.
Moreover, the fluorination of another carbon allotrope, graphene, has been extensively studied~\cite{TOUHARA2000241, PADAMATA2022109964,doi:10.1021/nn406333f, https://doi.org/10.1002/adma.202101665, https://doi.org/10.1002/advs.201500413}.
Therefore, we believe that our band engineering scheme can be realized in experiments so that fluorinated BPN could be an ideal platform to study intriguing phenomena from various types of Dirac dispersions and flat bands.

\begin{acknowledgements}
H.L and J.W.R are supported by the National Research Foundation of Korea (NRF) Grant
funded by the Korean government (MSIT) (Grant no.
2021R1A5A1032996). J.W.R is supported by the National Research Foundation of Korea (NRF) Grant
funded by the Korean government (MSIT) (Grant no.
2021R1A2C1010572 and
2022M3H3A106307411) and the Ministry of Education(Grant no. RS-2023-00285390). S.K is supported by the National Research Foundation of Korea (NRF) grant funded by the Korea government (Grant No.~NRF-2022R1F1A1074670). S.K.S acknowledges the support by the National Research Foundation of Korea(NRF) grant funded by the Korea government(MSIT) (2022R1A5A8033794).
\end{acknowledgements}

%

\end{document}